\documentclass[11pt,a4paper]{article}

\usepackage{jheppub} 
\usepackage{amssymb,amsmath}
\usepackage{subcaption}
\graphicspath{{./Fig/}}

\usepackage{graphicx}
\usepackage{color}

\renewcommand{\Re}{{\rm Re}}

\renewcommand{\Im}{{\rm Im}}

\newcommand{\SUN}[1]{{\rm SU}(N)}

\newcommand{\bbR}{{\mathbb R}}
\newcommand{\bbC}{{\mathbb C}}

\begin{document}

\title{
Gauge cooling for the singular-drift problem in the complex Langevin method
--- a test in Random Matrix Theory for finite density QCD
}

\author[a]{Keitaro Nagata,}
\author[a,b]{Jun Nishimura}
\author[a,c]{and Shinji Shimasaki}

\affiliation[a]{KEK Theory Center,
High Energy Accelerator Research Organization, Tsukuba 305-0801, Japan}
\affiliation[b]{Department of Particle and Nuclear Physics, 
School of High Energy Accelerator Science,\\
Graduate University for Advanced Studies (SOKENDAI), 
Tsukuba 305-0801, Japan}
\affiliation[c]{Research and Education Center for Natural Sciences, Keio University,\\
Hiyoshi 4-1-1, Yokohama, Kanagawa 223-8521, Japan}

\emailAdd{knagata@post.kek.jp}
\emailAdd{jnishi@post.kek.jp}
\emailAdd{simasaki@post.kek.jp}
\note{KEK-CP-322, KEK-TH-1854}

\date{\today
%,Since Jan .. 
}

\abstract{
Recently, the complex Langevin method has been applied successfully 
to finite density QCD either in the deconfinement phase or in the
heavy dense limit with the aid of a new technique called the gauge cooling.
In the confinement phase with light quarks, however,
convergence to wrong limits occurs due to 
the singularity in the drift term
caused by small eigenvalues of the Dirac operator including the mass term.
%Based on new insights into this problem, 
We propose that this singular-drift problem should also be 
overcome by the gauge cooling with different criteria for choosing
the complexified gauge transformation.
The idea is tested in chiral Random Matrix Theory for finite density QCD, 
where exact results are reproduced at zero temperature with light quarks.
It is shown that the gauge cooling indeed changes drastically
the eigenvalue distribution of the Dirac operator
%for configurations generated 
measured during the Langevin process.
Despite its non-holomorphic nature,
this eigenvalue distribution
% of the Dirac operator
has a universal diverging behavior at the origin
in the chiral limit
%fixed by the chiral condensate through 
due to 
a generalized Banks-Casher relation as we confirm explicitly.
%% We also derive and confirm a generalized Banks-Casher relation between
%% the chiral condensate and the eigenvalue distribution,
%% which will be useful in applications to QCD.
%%
%the latter being a non-holomorphic quantity, which does not have 
%direct connection to the corresponding quantity in the original path integral.
%
%We confirm this relation explicitly in the chiral Random Matrix Theory.
%
%% We also establish a generalized Banks-Casher relation, which 
%% relates the chiral condensate and 
%% the eigenvalue distribution of the Dirac operator
%% obtained by averaging over configurations generated during the 
%% Langevin process.
}

\maketitle

%%%%%%%%%%%%%%%%%%%%%%%%%%%%%%%%%%%%%%%%%%%%%%%%%%%
\section{Introduction}
%%%%%%%%%%%%%%%%%%%%%%%%%%%%%%%%%%%%%%%%%%%%%%%%%%%

Investigating the QCD phase diagram at finite temperature and 
finite density
is an important subject in theoretical physics
since it will provide us with microscopic understanding 
of various types of matter including nuclear matter. 
The standard Monte Carlo simulation, however, cannot be applied
due to the sign problem, which is caused by the complex-valued fermion determinant 
in the presence of the quark chemical potential. 
The sign problem is also of general interest since it appears 
in many important cases including investigations of
Chern-Simons gauge theory, supersymmetric gauge theories
and the real-time dynamics of quantum systems.
While there are several 
attempts within the framework of conventional Monte Carlo 
simulation such as reweighting, Taylor expansion, 
analytic continuation from the imaginary chemical potential
and the canonical approach,
they all suffer from increasing uncertainties 
with the increasing chemical potential.
(See refs.~\cite{Muroya:2003qs,deForcrand:2010ys} for reviews.) 
Recently, there was remarkable progress in a new type of approach
based on complexifying the dynamical variables.
In particular, the complex Langevin 
method (CLM) \cite{Parisi:1984cs,Klauder:1983sp}
was not only developed theoretically \cite{Aarts:2009dg,Aarts:2009uq,Aarts:2011ax,%
Seiler:2012wz,Nishimura:2015pba,Nagata:2015uga}
but also shown to work practically 
in various interesting cases \cite{Aarts:2013uxa,%
Sexty:2013ica,Fodor:2015doa,Sinclair:2015kva,Ichihara:2016uld}. 
As a closely related method, Monte Carlo simulation on 
the Lefshetz thimble \cite{Cristoforetti:2012su,Fujii:2013sra,DiRenzo:2015foa,%
Tanizaki:2015rda,Fujii:2015vha,Alexandru:2015xva,Scorzato:2015qts}
has also been studied intensively.
Comparison of the two methods is considered useful
in deepening our understanding in this approach
and in improving it \cite{Aarts:2014nxa,Tsutsui:2015tua,Hayata:2015lzj}.

The CLM \cite{Parisi:1984cs,Klauder:1983sp}
is an attempt to solve quantum systems with 
a complex-valued action using the idea of 
stochastic quantization based on the Langevin equation \cite{Parisi:1980ys}.
Since the stochastic quantization does not rely on 
the probability interpretation of the Boltzmann weight,
the method has a chance to be applicable to the case of complex action. 
In the case of real action, the Langevin equation is shown
to converge to the correct limit
%as can be shown by using
%owing to 
using the Fokker-Planck equation (See ref.~\cite{Damgaard:1987rr}, for instance.).
In the case of complex action, however,
there are some subtleties in the proof of the correct convergence.
Indeed, the CLM works well 
in some cases but fails 
in the other cases~\cite{Makino:2015ooa,Bloch:2015coa,Yamamoto:2015ura}.

In refs.~\cite{Aarts:2009uq,Aarts:2011ax},
%% In ref.~\cite{Aarts:2009uq}, 
%% %Aarts et al.\  clarified the conditions for the CLM to work.
the conditions for the CLM to work have been clarified.
In solving the Langevin equation for the complex action,
dynamical variables are necessarily complexified 
due to the complex drift term derived from the action.
It is important here that the observable as well as the drift term
is extended to complexified variables in a holomorphic fashion.
The expectation value of the observable 
is defined with the probability distribution
of the complexified variables.
%which also extends to the imaginary direction of variables.
%In ref.~\cite{Aarts:2009uq}, 
%% %Aarts et al.\  clarified the conditions for the CLM to work.
Then, under certain conditions, the expectation value thus defined
is shown to be equal to the expectation value defined in the original theory
for real variables with the complex weight. 
%They also pointed out possible causes of failure in this method.
% in this ses in which the argument fails.

One of the problems that can arise in this method is the runaway problem, 
which refers to the instability of the simulation.
This problem, however, was shown to be avoided
by using an adaptive step-size (rather than a fixed one)
for the discretized Langevin time \cite{Aarts:2009dg}.
A more serious problem is the convergence to wrong limits.
In ref.~\cite{Aarts:2009uq,Aarts:2011ax},
one of the causes of this problem
was identified to be the insufficient falloff of 
the probability distribution in the imaginary direction,
%which invalidates the aforementioned argument for justification.
which causes the violation of the aforementioned conditions.
We call it the excursion problem in this paper
since it occurs when the simulation makes a long excursion 
into the deeply imaginary regime.
%and/or the observable takes large absolute values in that regime.
%
%the strong growth of the observables in the imaginary direction, 

In order to cure the excursion problem,
ref.~\cite{Seiler:2012wz} proposed the gauge cooling,
which enabled the application of the CLM to certain parameter regions of QCD
such as the heavy dense limit \cite{Seiler:2012wz,Aarts:2013uxa}
and the deconfined phase \cite{Sexty:2013ica,Fodor:2015doa}.
In our previous paper~\cite{Nagata:2015uga}, 
we provided explicit justification of the gauge cooling
based on the argument given in refs.~\cite{Aarts:2009uq,Aarts:2011ax}.
We wrote down
how the probability distribution of the complexified variables 
evolves in the Langevin time 
under the influence of the gauge cooling procedure.
Then the corresponding Fokker-Planck equation
for the complex weight was shown
%that can be derived from the argument 
not to be affected by the gauge cooling 
as long as the observables are restricted to gauge invariant ones.

Whether the CLM works also in the confined phase 
with light quarks is still an open question, though.
This issue was addressed using chiral Random Matrix Theory (cRMT)
for finite density QCD at zero temperature \cite{Bloch:2012bh,Osborn:2004rf},
where a straightforward application of the CLM
led to wrong convergence when the quark mass becomes small~\cite{Mollgaard:2013qra}.
It was speculated that the problem is caused by 
the branch cut
%in the implimentation 
associated with logarithmic singularity in the 
action, which appears
from the fermion determinant \cite{Mollgaard:2013qra,Mollgaard:2014mga,Greensite:2014cxa}.
It was also pointed out~\cite{Seiler:2014sign} that 
the singular drift term derived from the logarithmic action
spoils the holomorphy, which is crucial in the CLM.
Recently, two of us (J.~N. and S.~S.) showed that 
the problem is not restricted to
the case with logarithmic singularities in the action 
but it occurs also in the case with higher order singularities \cite{Nishimura:2015pba}.
There it was also realized that the problem occurs in general when the probability
distribution of the complexified variables is nonnegligible for configurations
close to the singularity of the drift term.
%\footnote{It is important that
%the problem is not caused by the frequent crossing of the branch cut.}
Therefore, we refer to this problem as the singular-drift problem.

According to the new insights gained above,
the singular-drift problem is similar to the excursion problem
in that they are both related to the property of the
probability distribution of the complexified variables.
% plays an important role.
%In the forth-coming paper \cite{NNS},
Therefore, it is conceivable that
the gauge cooling, which is useful in overcoming the excursion
problem, is also useful in overcoming the singular-drift problem
if we choose the complexified gauge transformation used in the 
gauge cooling procedure appropriately.

In this paper, we test this idea in 
the cRMT, in which the singular-drift problem was 
shown to occur for light quarks \cite{Mollgaard:2013qra}.
While the cRMT is not a gauge theory, the action and the observables have
U($N$) symmetries,
which become GL($N$,$\bbC$) symmetries upon complexifying the dynamical variables.
%Analogously to the case of gauge theories,
These complexified symmetries provide us with large enough freedom 
to modify the probability distribution
of the complexified variables in such a way that the singularity is avoided.
In the original gauge cooling for the excursion problem,
the complexified gauge transformation is chosen
to reduce the so-called unitarity norm,
which provides an estimate on the deviation of 
link variables from SU(3) matrices. 
On the other hand, the singular-drift problem occurs when 
the Dirac operator (including the mass term)
defined for the complexified dynamical variables
has eigenvalues close to zero.
%with small absolute values
%close to zero,
Therefore, we introduce new types of norm, which are sensitive to 
the properties of the Dirac operator.
%the eigenvalue distribution of the Dirac operator.
Using the gauge cooling with the new types of norm, we show that
the eigenvalue distribution of the Dirac operator measured during
the simulation is indeed modified in such a way that the singularity
of the drift term is avoided unless the quark mass becomes too small.
Thus, exact results are nicely reproduced for quark mass, which is much
smaller than that achieved by the CLM without gauge cooling.

%Some comments on our results are in order.
%First, 
In fact,
the eigenvalue distribution of the Dirac operator for the complexified variables
is invariant under GL($N$,$\bbC$) transformations.
However, the Langevin time evolution itself is affected nontrivially
by the gauge cooling procedure due to the noise term, which
does not transform covariantly under GL($N$,$\bbC$) transformations.
As a result, the eigenvalue distribution measured during the Langevin process
is modified by the gauge cooling.
%for configurations obtained 
Interestingly, the eigenvalue distribution obtained in the thermal 
equilibrium of the Langevin process
depends on the choice of norm for the gauge cooling 
even in the cases where the exact results are reproduced.
This is possible because
the eigenvalue distribution under discussion
is not a holomorphic quantity and hence 
has no direct connection to the corresponding quantity 
in the original path integral with the complex weight.

On the other hand, the chiral condensate, which is a holomorphic quantity,
can be expressed in terms of the eigenvalue distribution \cite{Splittorff:2014zca}.
This implies that the non-universal eigenvalue distribution still
has a universal property, which does not depend on the norm 
used for gauge cooling as long as the CLM is working.
%We clarify this point by deriving 
In particular, we derive
a generalized Banks-Casher relation,
which implies that the eigenvalue distribution has a universal
diverging behavior at the origin in the chiral limit,
and show that it is indeed satisfied in the cRMT.
As the original Banks-Casher relation is useful in QCD 
at zero chemical potential,
the generalized version is expected to be useful
in applying the CLM to QCD at finite density.

%% In ref.~\cite{Mollgaard:2014mga}, 
%% it was shown that the singular drift problem in the cRMT 
%% can be overcome completely by 
%% using the polar coordinates for each complex element of the matrices.
%% %
%% a suitable choice of a  coordinate solves this problem
%% for the chiral random matrix theory. 
%% However, it is unclear that such a choice of coordinate 
%% helps to solve the problem in QCD. 
%% The method introduced in this work can be applied to a wide class of theories. 
%% Then, the second generalization is that the gauge cooling is not limited to QCD, 
%% but can be used for other theories, if there are sufficiently large symmetries to 
%% transform all the dynamical variables.
%% Although we test the new method 
%% for the chiral random matrix theory, the method we propose can be 
%% extended to QCD in a straightforward manner. 

The rest of this paper is organized as follows. 
In section \ref{sec:framework}, 
we review some basic features of the CLM and
its application to the cRMT.
In section \ref{sec:cooling}, we discuss
the singular-drift problem, which occurs
in the CLM when the quark mass is small.
We also discuss how this problem can be overcome by
the gauge cooling using new types of norm.
%small eigenvalues of the Dirac operator, which 
%appear during the Langevin process 
%explain how the CLM fails due to 
%small eigenvalues of the Dirac operator, which 
%appear during the Langevin process when the quark mass is small.
%We also discuss how this problem can be overcome by
%the gauge cooling using new types of norm.
In section \ref{sec:result}, we present our results and show that 
the exact results of the cRMT can be reproduced for quark mass
much smaller than that achievable without gauge cooling.
We also present the eigenvalue distribution of the Dirac operator
measured during the Langevin process, and show that the singularity
of the drift term is avoided in differently ways depending 
on the norm adopted.
In section \ref{sec:discussion}, we derive
a generalized Banks-Casher relation, which connects 
the chiral condensate in the chiral limit
to the asymptotic behavior of the
eigenvalue distribution of the Dirac operator at the origin. 
% measured in the CLM.
We show that the relation is indeed satisfied in the cRMT,
and discuss its implication in the context of this work.
Section \ref{sec:summary} is devoted to a summary and discussions. 
Some preliminary results of this work have already been presented
in a proceeding contribution \cite{Nagata:2015ijn}.

%%%%%%%%%%%%%%%%%%%%%%%%%%%%%%%%%%%%%%%%%%%%%%%%%%%%%%%%%%%%%%%%%%%%%%%%%%%%%%%%%%%%%%%%%%
\section{The method and the model}
\label{sec:framework}
%%%%%%%%%%%%%%%%%%%%%%%%%%%%%%%%%%%%%%%%%%%%%%%%%%%%%%%%%%%%%%%%%%%%%%%%%%%%%%%%%%%%%%%%%%

In this section, we briefly review some basic features of 
the CLM and discuss its application to the cRMT.

%%%%%%%%%%%%%%%%%%%%%%%%%%%%%%%%%%%%%%%%%%%%%%%%%%
\subsection{the complex Langevin method (CLM)}
%%%%%%%%%%%%%%%%%%%%%%%%%%%%%%%%%%%%%%%%%%%%%%%%%%

Let us consider a system of $n$ real variables $x_i,\;(i=1,\cdots,n)$ 
defined by the partition function
\begin{align}
Z=\int \prod_{i=1}^{n} dx_i \; e^{-S(\{ x_i\})} \ .
\end{align}
The basic idea of the stochastic quantization is 
to investigate this system using the Langevin equation \cite{Parisi:1980ys}
\begin{align}
\frac{ d x_i(\tau)}{d \tau} = 
- \frac{\partial S(\{ x_i(\tau) \})}{\partial x_i} + \eta_i(\tau) \ , 
\label{Eq:2015Jul27eq1}
\end{align}
where $\tau$ is a fictitious time dubbed the Langevin time, 
and $\eta_i(\tau)$ is a Gaussian random noise normalized by 
$\langle \eta_i (\tau) \eta_j(\tau') \rangle = 2 \delta_{ij} \delta(\tau-\tau')$.
When the action $S(\{ x_i\})$ is real, one can show that
the average of an observable
over sufficiently long Langevin time agrees with the expectation value
of the observable in the original path integral \cite{Damgaard:1987rr}; namely 
\begin{align}
\langle {\cal O} \rangle = 
\lim_{T\to \infty} \frac{1}{T} \int_{\tau_0}^{\tau_0+T} d\tau \;O(\{ x_i(\tau) \}) \ , 
\label{Eq:2015Jul27eq2}
\end{align}
where $\tau_0$ represents the Langevin time necessary for thermalization
and $T$ represents the total Langevin time used 
for averaging.

Since the method does not rely on the probability interpretation of 
the factor $e^{-S(\{ x_i\})}$, it has a chance to be generalized to 
the case of complex action \cite{Parisi:1984cs,Klauder:1983sp}.
When the action is complex, however, the 
drift term $- \frac{\partial S}{\partial x_i}$ 
in (\ref{Eq:2015Jul27eq1}) becomes complex, and 
one has to allow the dynamical variables to take complex values as
$x_i \in \mathbb{R} \to z_i \in \mathbb{C}$ during the Langevin process.
The Langevin equation (\ref{Eq:2015Jul27eq1}) should then be replaced by
\begin{align}
\frac{ d z_i(\tau)}{d \tau} = 
- \frac{\partial S(\{ z_i(\tau) \})}{\partial z_i} + \eta_i(\tau) \ , 
\label{cle}
\end{align}
where the action $S$ is now
regarded as a holomorphic function of $z_i$ 
defined through analytic continuation.
The expectation value can be calculated by (\ref{Eq:2015Jul27eq2})
with $O(\{ x_i (\tau)\})$ being replaced by $O(\{ z_i (\tau)\})$,
which is also defined through analytic continuation.
Unlike the case of real action, there are certain conditions that should be
met for the method to work in the case of 
complex action \cite{Aarts:2009uq,Aarts:2011ax}.
In particular, it is required that 
the probability distribution of the complexified variables
should have a fast fall-off in the imaginary direction. 
For this reason, it is considered better to leave the noise term
in (\ref{cle}) real \cite{Aarts:2009uq}
although it can be generalized to complex values in principle.

%% However, the singularity in drift terms violates those conditions. 
%% In this work, we consider the real noise in the complex Langevin dynamics. 

%%%%%%%%%%%%%%%%%%%%%%%%%%%%%%%%%%%%%%%%%%%%%%%%%%
\subsection{chiral Random Matrix Theory (cRMT)}
\label{sec:cRMT}
%%%%%%%%%%%%%%%%%%%%%%%%%%%%%%%%%%%%%%%%%%%%%%%%%%

We consider the cRMT for $N_{\rm f}$ quarks with degenerate mass $m>0$ 
at zero temperature and finite chemical potential $\mu$ \cite{Bloch:2012bh,Osborn:2004rf}.
%introduced in \cite{Osborn:2004rf}.
The partition function is defined by \cite{Bloch:2012bh}
\begin{align}
Z &= \int d\Phi_1d\Phi_2 \,  [\det (D+m)]^{N_{\rm f}} e^{-S_{\rm b}}  \ ,
\label{crmt}
\end{align}
%% \begin{align}
%% Z &= \int d\Phi_1d\Phi_2 \, [\det (D+m)]^{N_{\rm f}} e^{-S_{\rm b}}  \ , 
%% \label{crmt}
%% \end{align}
%which was introduced in ~\cite{Bloch:2012bh} as a variation of the 
%partition function introduced in ~\cite{Osborn:2004rf}.
where $\Phi_{k} \ (k=1,2)$ are general $N\times (N+\nu)$ complex 
matrices.\footnote{The model (\ref{crmt})
is equivalent to the one investigated in ref.~\cite{Mollgaard:2013qra}
as one can show by a simple change of variables~\cite{Bloch:2012bh}.
% via the following relation, 
%\begin{align}
%\Phi \to \Phi_1, \Psi \to - \Phi_2 ^\dagger. 
%\end{align}
}
The integer $\nu$ represents the topological index, which gives 
the number of exact zero eigenvalues of the Dirac operator $D$ 
given by 
\begin{align}
D&=\left( \begin{matrix}
0 & X \\
Y & 0
\end{matrix} \right)  \ , 
\label{eq:Dirac}
\\
X&= e^{\mu} \Phi_1 + e^{-\mu} \Phi_2 \ , 
\label{Eq:2015Sep01eq1}\\
Y&=-e^{-\mu} \Psi_1 - e^\mu \Psi_2 \  ,
\label{Eq:2015Sep01eq2}
\end{align}
where we have defined $(N+\nu)\times N$ matrices $\Psi_k \ (k=1,2)$ by
\begin{align}
\Psi_k=(\Phi_k)^\dagger 
\label{hc-constraint}
\end{align}
for later convenience.
The bosonic action $S_{\rm b}$ in (\ref{crmt}) is given by 
\begin{align}
S_{\rm b}&=2N \sum_{k=1}^{2}{\rm Tr} (\Psi_k \Phi_k) \ .
\label{Sb} 
\end{align}
The effective action of the system can be written as
\begin{align}
%S_{\rm eff}&= S_{\rm b} - N_{\rm f} \ln \det (m^2 - XY) \ .
S_{\rm eff}&= S_{\rm b} - N_{\rm f} \ln \det (D+m) \ .
\label{Seff} 
\end{align}
%weight $w (\Phi_1 ,  \Phi_2)$ 
%\footnote{
Strictly speaking,
the logarithm in (\ref{Seff}) has an ambiguity due to 
the branch cut \cite{Mollgaard:2013qra,Mollgaard:2014mga}.
This is not an issue, however, since the CLM can be formulated
in terms of the weight 
%$w=e^{-S}$ 
$w=[\det (D+m)]^{N_{\rm f}} e^{-S_{\rm b}}$
without ever having to refer to
the effective action (\ref{Seff}) 
as was pointed out in ref.~\cite{Nishimura:2015pba}.
With that in mind, we keep on using the 
effective action (\ref{Seff}) just for simplicity of terminology.

As observables in this model, one can consider
the chiral condensate $\Sigma$ and the baryon number density $n_B$ defined by
\begin{align}
\Sigma = \frac{1}{N} \frac{\partial}{\partial m} \log Z  \ , \quad \quad
n_B = \frac{1}{N} \frac{\partial}{\partial \mu} \log Z \ .
%= \frac{\bar{q}q}{N} &= \frac{2 m N_{\rm f}}{N} {\rm Tr} [ (m^2-XY)^{-1}],  \\
%\frac{q^\dagger q}{N} &= \frac{N_{\rm f}}{N} {\rm Tr} [ (m^2-XY)^{-1} \frac{\partial}{\partial \mu} (m^2-XY)].
\label{chiral}
\end{align}
The partition function of the cRMT can be calculated analytically
using the orthogonal polynomial method \cite{Osborn:2004rf}.
It turns out that the partition function is independent of the chemical potential 
$\mu$, and hence the baryon number density is exactly zero
and the chiral condensate has no $\mu$ dependence.
(See, e.g., ref.~\cite{Mollgaard:2013qra} for an explicit expression
of the chiral condensate $\Sigma$ suitable for numerical evaluation.)

In the gauge cooling, symmetries of the system play a crucial role.
Let us consider a transformation
%$\mathrm{U}(N)\times \mathrm{U}(N+\nu)$ transformation,
\begin{align}
\Phi_k \to \Phi_k' = g \Phi_k h^{-1},\quad
\Psi_k \to \Psi_k' = h \Psi_k g^{-1}, \quad (k=1,2) \ ,
\label{Eq:2015Apr14eq1}
\end{align}
which leaves the bosonic action invariant.
In order for (\ref{hc-constraint}) to be satisfied 
for the transformed configuration,\footnote{Upon complexifying the variables
in the CLM, the constraint (\ref{hc-constraint}) is disregarded,
and hence the symmetry enhances to (\ref{Eq:2015Apr14eq1}) with
$g\in \mathrm{GL}(N)$ and $h\in \mathrm{GL}(N+\nu)$.
In view of this, we use $g^{-1}$ and $h^{-1}$ instead of
$g^{\dagger}$ and $h^{\dagger}$.
}
we need to have $g\in \mathrm{U}(N)$ and $h\in \mathrm{U}(N+\nu)$.
The matrices $X$ and $Y$ 
in (\ref{Eq:2015Sep01eq1}) and (\ref{Eq:2015Sep01eq2})
transform in the same way as 
$\Phi_k$ and $\Psi_k$, respectively, and hence the Dirac operator $D$
defined by (\ref{eq:Dirac}) transforms as
\begin{align}
\left( \begin{matrix}
0 & X \\
Y & 0
\end{matrix} \right)  
\to
\left( \begin{matrix}
g & ~ \\
~  & h
\end{matrix} \right)  
\left( \begin{matrix}
0 & X \\
Y & 0
\end{matrix} \right)  
\left( \begin{matrix}
g^{-1} & ~ \\
~ & h^{-1}
\end{matrix} \right)   \ .
\label{D-transform}
\end{align}
Therefore, 
both the effective action (\ref{Seff}) and the observables (\ref{chiral})
are invariant under the transformation (\ref{Eq:2015Apr14eq1}).

Note that the Dirac operator $D$ satisfies
%anticommutes with the chiral matrix
%Note that the eigenvalues of $D$ appear in $\pm$-pairs since $D$ anticommutes with the chiral matrix 
\begin{align}
D \gamma_5  & = - \gamma_5 D  \ ,
\label{anticommuting}
\\
 \gamma_5 &=\begin{pmatrix} {\bf 1}_{N} & 0 \\ 0 & -{\bf 1}_{N+\nu} \end{pmatrix}
\label{gamma5-def}
\end{align}
for any $\mu$,
%; i.e., $D \gamma_5 = - \gamma_5 D $, 
which implies
that the nonzero eigenvalues of $D$ appear in pairs with opposite signs.
Furthermore, when $\mu=0$, the Dirac operator $D$ is anti-Hermitian 
$D^\dagger = - D$ and hence its eigenvalues are purely imaginary.
Thus, one can show that $\mathrm{det}(D+m)$ is real positive in this case.
On the other hand, when $\mu\neq 0$, $D$ is no longer anti-Hermitian,
and its eigenvalues take general complex values.
The determinant $\mathrm{det}(D+m)$ 
%and hence the weight $w (\Phi_1 ,  \Phi_2)$ in (\ref{crmt}) 
becomes complex in general, 
which causes the sign problem.

%%%%%%%%%%%%%%%%%%%%%%%%%%%%%%%%%%%%%%%%%%%%%%%%%%
\subsection{application of the CLM to the cRMT}
\label{sec:clm-cRMT}
%%%%%%%%%%%%%%%%%%%%%%%%%%%%%%%%%%%%%%%%%%%%%%%%%%

In order to apply the CLM to this system, 
we introduce real variables $(x_k)_{ij}$ and $(y_k)_{ij}$ through 
\begin{align}
(\Phi_k)_{ij} & = (x_k)_{ij} + i (y_k)_{ij} \ , \\ 
(\Psi_k)_{ji} & = (x_k)_{ij} - i (y_k)_{ij} \ , 
\label{xy-def}
\end{align}
taking account of (\ref{hc-constraint}).
The effective action (\ref{Seff})
and the observables (\ref{chiral}) are functions of these real variables.
Then we complexify the variables as
$(x_k)_{ij} \to (z_k)_{ij} \in \mathbb{C}$ and $(y_k)_{ij} \to (w_k)_{ij} \in \mathbb{C}$,
redefining the action and the observables as holomorphic functions of
$(z_k)_{ij}$ and $(w_k)_{ij}$ by analytic continuation.
The complex Langevin equation (\ref{Eq:2015Jul27eq1}) can be derived
in the present case in a straightforward manner \cite{Mollgaard:2013qra}.
After complexifying the variables,
the effective action and the observables may be regarded as
holomorphic functions of $({\Phi}_k)_{ij} $ and $({\Psi}_k)_{ji}$,
where the constraint (\ref{hc-constraint}) is no longer imposed.
Therefore, they are now invariant under (\ref{Eq:2015Apr14eq1})
with $g\in \mathrm{GL}(N)$ and $h\in \mathrm{GL}(N+\nu)$;
namely the symmetry is doubly enhanced.
We use this enhanced symmetry for the gauge cooling.

%% The symmetry (\ref{Eq:2015Apr14eq1}) of the action and the observables
%% are enhanced after complexifying the variables.
%% This can be seen by defining independent complex matrices
%% \begin{align}
%% (\tilde{\Phi}_k)_{ij} & = (z_k)_{ij} + i (w_k)_{ij} \ , \\ 
%% (\tilde{\Psi}_k)_{ji} & = (z_k)_{ij} - i (w_k)_{ij} \ .
%% \label{xy-def-complexified}
%% \end{align}
%% The effective action and the observables can be regarded
%% as holomorphic functions of $(\tilde{\Phi}_k)_{ij} $ and $(\tilde{\Psi}_k)_{ji}$,
%% which can be obtained 
%% from (\ref{Seff}) and (\ref{chiral})
%% by replacing 
%% $\Phi_k$ and $\Psi_k$ with $\tilde{\Phi}_k$ and $\tilde{\Psi}_k$, respectively.
%% Note here that there is no constraint such as (\ref{hc-constraint})
%% between $\tilde{\Phi}_k$ and $\tilde{\Psi}_k$.
%% Because of this, the effective action and the observables are now invariant 
%% under
%% \begin{align}
%% \tilde{\Phi}_k \to \tilde{\Phi}_k' = g \tilde{\Phi}_k h^{-1},\quad
%% \tilde{\Psi}_k \to \tilde{\Psi}_k' = h \tilde{\Psi}_k g^{-1}, \quad (k=1,2) \ ,
%% \label{Eq:2015Apr14eq1-complexified}
%% \end{align}
%% where $g\in \mathrm{GL}(N)$ and $h\in \mathrm{GL}(N+\nu)$.
%% Thus the symmetry is doubly enhanced.

%%%%%%%%%%%%%%%%%%%%%%%%%%%%%%%%%%%%%%%%%%%%%%%%%%%%%%%%%%%%%%%%%%%%%%%%
\section{Gauge cooling for the singular-drift problem}
\label{sec:cooling}
%%%%%%%%%%%%%%%%%%%%%%%%%%%%%%%%%%%%%%%%%%%%%%%%%%%%%%%%%%%%%%%%%%%%%%%%

%the Dirac operator including the mass term
%propose to cure this problem by the gauge cooling with new types of norm.
%
%In applying the CLM to the cRMT,
%this problem occurs when the eigenvalue distribution of $D+m$ near the origin
%becomes significant.
The gauge cooling was originally proposed 
to cure the excursion problem 
in the CLM for finite density QCD~\cite{Seiler:2012wz,Sexty:2013ica}. 
The basic idea is to reduce the unitarity norm, 
which measures the deviation of the link variables from SU($3$) matrices,
by making a complexified gauge transformation after each Langevin step
when one solves the discretized complex Langevin equation.
While the gauge cooling procedure changes the Langevin dynamics nontrivially,
the expectation values of gauge invariant observables are expected to be 
obtained correctly once the problem is cured.
Recently, this has been proved explicitly \cite{Nagata:2015uga} 
by extending the argument for justification of the CLM 
\cite{Aarts:2009uq,Aarts:2011ax} to the case with the gauge cooling procedure.

Let us recall here that the drift term involves
${\rm Tr}[ (D+m)^{-1} \partial (D+m)]$, which becomes singular
when the operator $D+m$ has zero eigenvalues.
%As we mentioned in the previous section, 
For $\mu=0$,
the CLM reduces to the real Langevin simulation,
and $D$ becomes an anti-Hermitian matrix, whose
eigenvalues are purely imaginary.
In this case, all the eigenvalues of $D+m$ lie on a straight line parallel
to the imaginary axis on the complex plane.
As $\mu$ is increased, the eigenvalue distribution is broadened
in the direction of the real axis, and the singular-drift problem
occurs for $\mu$ larger than some critical value depending 
on $m$ \cite{Mollgaard:2013qra}.
We propose to avoid this problem
by using the gauge cooling with
the complexified symmetry transformation chosen
in such a way that the eigenvalues of $D+m$ do not appear near the origin.

%%%%%%%%%%%%%%%%%%%%%%%%%%%%%%%%%%%%%%%%%%%%%%%%%%
\subsection{New types of norm for the gauge cooling}
%%%%%%%%%%%%%%%%%%%%%%%%%%%%%%%%%%%%%%%%%%%%%%%%%%

As we mentioned in section \ref{sec:clm-cRMT},
the symmetry (\ref{Eq:2015Apr14eq1})
of the effective action and the observables in the cRMT
enhances from $\mathrm{U}(N)\times \mathrm{U}(N+\nu)$ 
to $\mathrm{GL}(N,\mathbb{C})\times \mathrm{GL}(N+\nu,\mathbb{C})$
upon complexification of the dynamical variables.
We use this complexified symmetry to perform the gauge cooling.
The transformation (\ref{Eq:2015Apr14eq1}) to be made after each
Langevin step is determined by minimizing
some positive definite quantity ``norm'',
which is invariant under the $\mathrm{U}(N)\times \mathrm{U}(N+\nu)$ transformation
but not under $\mathrm{GL}(N,\mathbb{C})\times \mathrm{GL}(N+\nu,\mathbb{C})$ 
transformation.

First, let us consider the ``Hermiticity norm''
\begin{align}
\mathcal N_{\rm H} &= \frac{1}{N}\, {\rm tr} \sum_{k=1,2}
[ (\Psi_k-\Phi_k^\dagger)^\dagger (\Psi_k-\Phi_k^\dagger) ]  \\
&= \frac{4}{N} \sum_{k=1,2} \sum_{ij}
\Big[ \{ {\rm Im} (z_k)_{ij} \}^2  + \{ {\rm Im} (w_k)_{ij} \}^2  \Big] \ ,
\end{align}
which measures the violation of the relation (\ref{hc-constraint}).
Reducing this norm has an effect of keeping the complexified variables
closer to real values, and it may be regarded as an analog of 
the unitarity norm in QCD \cite{Seiler:2012wz,Sexty:2013ica}.
It turns out, however, that the gauge cooling with the Hermiticity norm 
does not help in curing the singular-drift problem.
This is understandable since the eigenvalue distribution
of $D+m$ is not directly related to the property (\ref{hc-constraint}).
In order to cure the singular-drift problem,
we therefore need to consider other types of norm,
whose reduction affects the eigenvalue distribution directly.

Here we propose two types of norm, 
which can be used for the gauge cooling to
cure the singular-drift problem.
The first one is given by
\begin{align}
\mathcal N_1 = \frac{1}{N}{\rm Tr}\left[(X+Y^\dagger)^\dagger(X+Y^\dagger)\right] \ ,
\label{normtype1}
\end{align}
where $X$ and $Y$ are defined 
by \eqref{Eq:2015Sep01eq1} and \eqref{Eq:2015Sep01eq2}, respectively,
and the dagger in $Y^\dagger$ implies taking the Hermitian conjugate of $Y$
after complexifying the dynamical variables.
Since this norm vanishes if and only if $D$ is anti-Hermitian,
it provides a measure of the deviation of $D$ from an anti-Hermitian matrix.
Reducing this norm has an effect of making the eigenvalue distribution 
of $D+m$ narrower in the real direction, and hence it is expected to
cure the singular-drift problem.

The second one is defined by 
\begin{align}
\mathcal N_{2} = \sum_{a=1}^{n_{\rm ev}} e^{-\xi \alpha_a} \ ,
\label{normtype2}
\end{align}
where $\alpha_a$ are the real positive eigenvalues of $M^\dagger M$ 
with $M=D+m$ and $\xi$ is a real parameter.
Here again the dagger in $M^\dagger$ implies taking the Hermitian conjugate of $M$
after complexifying the dynamical variables.
The sum in (\ref{normtype2})
is taken over the $n_{\rm ev}$ smallest eigenvalues of $M^\dagger M$.
Reducing this norm suppresses the appearance of small $\alpha_a$
and tries to achieve $\alpha_a > 1/\xi$. 
Since $\alpha_a > 1/\xi$ implies $|\lambda_a|^2 > 1/\xi$,
where $\lambda_a$ are the eigenvalues of $M$,
%Since $|\lambda_i|^2 \ge \alpha_1$,
%where $\lambda_i$ are the eigenvalues of $M$,
%the eigenvalues of $M$, which we denote as $\lambda_i$,   ,s are 
the appearance of $\lambda_a$ close to zero is also suppressed.
%makes the small eigenvalues of $M$ more difficult to appear.

In actual simulations, the excursion problem and the singular-drift problem
may occur at the same time. 
In that case, we take a linear combination of 
the Hermiticity norm and one of the new types of norm as 
\begin{align}
\mathcal N_{i}(s) = s \mathcal N_{\rm H} + (1-s) \mathcal N_i 
\quad \quad \quad
\mbox{for~}i=1,2 \ ,
\label{Ntot}
\end{align}
where 
%$\mathcal N_i$ is either $\mathcal N_1$ or $\mathcal N_2$, 
%$i=1,2$ and 
$s$ $(0 \le s \le 1)$ is a tunable parameter.

Note that the eigenvalue distribution of the Dirac operator $D$
is invariant 
under $\mathrm{GL}(N,\mathbb{C})\times \mathrm{GL}(N+\nu,\mathbb{C})$.
However, the Langevin dynamics is modified 
nontrivially\footnote{The symmetry enhancement from
${\rm U}(N)\times {\rm U}(N+\nu)$ to 
$\mathrm{GL}(N,\mathbb{C})\times \mathrm{GL}(N+\nu,\mathbb{C})$
occurs for the action and the observables but not for the Langevin process
itself. Despite this fact, the use of gauge cooling in the CLM 
can be justified. See ref.~\cite{Nagata:2015uga} for explicit demonstration
based on the Fokker-Planck equation.}
since the noise term respects only ${\rm U}(N)\times {\rm U}(N+\nu)$.
As a result, the eigenvalue distribution
for the configuration obtained in the next Langevin step
is affected nontrivially by the gauge cooling 
even if one averages over the Gaussian noise.
%Performing the cooling procedure after each Langevin step, the configuration 
%has been forced to have such a distribution of the eigenvalues that the imposed 
%norms are appropriately suppressed.

%%%%%%%%%%%%%%%%%%%%%%%%%%%%%%%%%%%%%%%%%%%%%%%%%%
\subsection{Details of the gauge cooling procedure}
%%%%%%%%%%%%%%%%%%%%%%%%%%%%%%%%%%%%%%%%%%%%%%%%%%

Below we explain the gauge cooling procedure in more detail.
After each Langevin step, 
we perform the transformation (\ref{Eq:2015Apr14eq1}),
where the transformation matrices
$g\in \mathrm{GL}(N)$ and $h\in \mathrm{GL}(N+\nu)$
%$\mathrm{GL}(N,\mathbb{C})\times \mathrm{GL}(N,\mathbb{C})$ 
are determined in such a way that a given norm is reduced efficiently.
Following the original proposal \cite{Seiler:2012wz}, we
first calculate the gradient of the norm with respect to the complexified
transformation. This can be done by considering the infinitesimal 
transformation
\begin{align}
g = 1 + \epsilon_a \lambda_a \ , 
\quad \quad
h = 1 + \delta_a \rho_a \ , 
\label{infinitesimal-tr}
\end{align}
where $\lambda_a$ and $\rho_a$ are the basis of $N\times N$ and
$(N+\nu) \times (N+\nu)$ Hermitian matrices, respectively,
normalized as 
${\rm tr} (\lambda_a \lambda_b)={\rm tr} (\rho_a \rho_b)= \delta_{ab}$.
Note that
(\ref{infinitesimal-tr}) corresponds to
an infinitesimal $\mathrm{U}(N) \times \mathrm{U}(N+\nu)$ transformation
when $\epsilon_a$ and $\delta_a$ are purely imaginary.
Considering that the norm is invariant under 
$\mathrm{U}(N) \times \mathrm{U}(N+\nu)$,
we assume that $\epsilon_a$ and $\delta_a$ are real in what follows.

Under the infinitesimal transformation (\ref{infinitesimal-tr}),
$\Phi_k$ and $\Psi_k$ transform as
\begin{align}
\delta \Phi_k  = \epsilon_a \lambda_a \Phi_k - \delta_a \Phi_k \rho_a \ , \\
\delta \Psi_k  = \delta_a \rho_a \Psi_k - \epsilon_a \Psi_k \lambda_a  \ .
\label{infinitesimal-tr-Phi-Psi}
\end{align}
We can calculate the change of the norm as
\begin{align}
\delta \mathcal N = \epsilon_a G_a + \delta_a H_a \ ,
\label{infinitesimal-tr-norm}
\end{align}
neglecting higher order terms in $\epsilon_a$ and $\delta_a$,
where $G_a , H_a \in \bbR$ represent the gradient of the norm.
%% can be obtained explicitly once the norm is given.
%% The gradient of the norm is given by
%% \begin{align}
%% 
%% \label{gradient}
%% \end{align}
We may reduce the norm most efficiently 
by choosing $(\epsilon_a , \delta_a)  \propto - (G_a , H_a)$
at the linearized level (\ref{infinitesimal-tr}).
As a finite transformation, we consider\footnote{In order to obtain $g$
in (\ref{finitesimal-tr-g}) numerically, 
we diagonalize the Hermitian matrix $G=U \Lambda U^\dagger$
with a unitary matrix $U$, and calculate $g= U e^{-\alpha \Lambda} U^\dagger$.
The matrix $h$ in (\ref{finitesimal-tr-h}) is obtained similarly.
}
\begin{align}
\label{finitesimal-tr-g}
g &= \exp (- \alpha G) \ , \quad \quad G = G_a \lambda_a \ , \\
h &= \exp (- \alpha H) \ , \quad \quad H = H_a \rho_a \ ,
\label{finitesimal-tr-h}
\end{align}
where the real parameter $\alpha$ is determined in such a way that
the norm for the transformed configuration of $\Phi_k$ and $\Psi_k$
is minimized. (In practice, this minimization is done approximately
because we can calculate the norm only for a finite number of $\alpha$.)
The above procedure is repeated until the norm is more or less
minimized with respect to
the $\mathrm{GL}(N,\mathbb{C})\times \mathrm{GL}(N+\nu,\mathbb{C})$ transformation.

%%%%%%%%%%%%%%%%%%%%%%%%%%%%%%%%%%%%%%%%%%%%%%%%%%%%%%%%%%%%%%%%%%%%%%%%
\section{Results of the CLM with or without gauge cooling}
\label{sec:result}
%%%%%%%%%%%%%%%%%%%%%%%%%%%%%%%%%%%%%%%%%%%%%%%%%%%%%%%%%%%%%%%%%%%%%%%%

In this section we present our results of the CLM for the cRMT.
The Langevin simulation is 
performed\footnote{Although we used a fixed step-size during the simulation
instead of an adaptive one \cite{Aarts:2009dg},
we did not encounter a runaway problem.
%This may be due to the fact that the Gaussian term
%in the bosonic action $S_{\rm b}$ induces
%a drift term, which attracts the configuration toward the origin.
}
with the step-size $\epsilon = 5\times 10^{-5}$.
We discard the first 20000 steps for thermalization;
i.e., $\tau_0=1$ in eq.~(\ref{Eq:2015Jul27eq2}).
Then we perform
$80000$ steps, which corresponds to $T=4$ in eq.~(\ref{Eq:2015Jul27eq2}),
during which we measure the observables every 200 steps.
After each Langevin step, we perform the gauge cooling, which amounts to
making a
$\mathrm{GL}(N,\mathbb{C})\times \mathrm{GL}(N+\nu,\mathbb{C})$
transformation (\ref{finitesimal-tr-g}) and (\ref{finitesimal-tr-h})
ten times as described in the previous section.
As for the parameter $s$ in eq.~(\ref{Ntot}),
we choose $s=0$ for the norm $\mathcal N_1(s)$ 
and $s=0.01$ for the norm $\mathcal N_2(s)$.
The parameters in the definition
(\ref{normtype2}) of the norm $\mathcal N_2$
are chosen as $\xi=300$ and $n_{\rm ev}=2$.

\begin{figure}[htbp] %%%%%%%%%%%%%%%%%%%%%%%%%%%%%%%%%%%%%%%%%%%%%%%%%
\includegraphics[width=7cm]{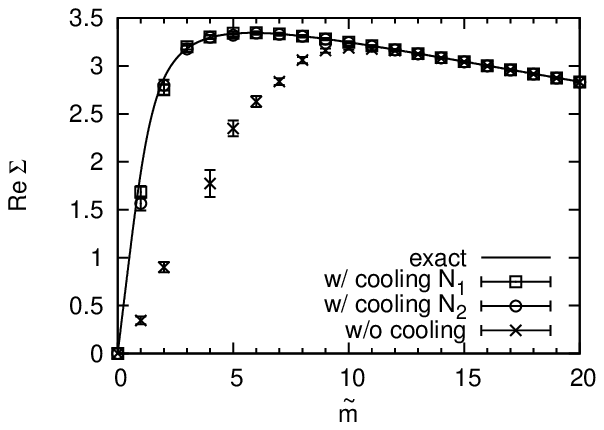}
\includegraphics[width=7cm]{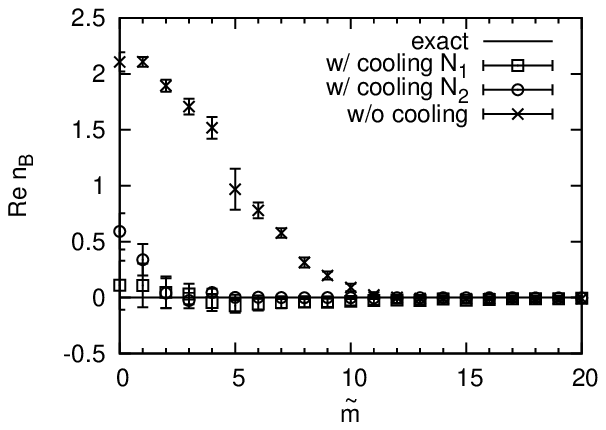}
\caption{The chiral condensate (Left) and the baryon number density (Right)
are plotted against $\tilde{m} \equiv m N$.
The results are obtained by the CLM for the cRMT with
$\nu=0$, $N_{\rm f}=2$, $N=30$ and $\tilde{\mu} \equiv \mu \sqrt{N} = 2$
with or without gauge cooling.
The solid lines represent the exact results for the cRMT.}
\label{Fig:2015Jul29Fig1}
\end{figure}%%%%%%%%%%%%%%%%%%%%%%%%%%%%%%%%%%%%%%%%%%%%%%%%%%%%%%%%%%

In Fig.~\ref{Fig:2015Jul29Fig1} we plot our results 
for the chiral condensate and the baryon number density
%of the CLM for the cRMT 
obtained with or without gauge cooling against $\tilde{m} \equiv m N$.
Here we set the parameters of the cRMT as
$\nu=0$, $N_{\rm f}=2$, $N=30$ and $\tilde{\mu} \equiv \mu \sqrt{N}  = 2$,
which were used in ref.~\cite{Mollgaard:2013qra} to reveal the problem
at small quark mass.\footnote{The cRMT (\ref{crmt})
is equivalent to the chiral perturbation theory 
(the low energy effective theory of QCD)
in the $\epsilon$-domain 
in the large-$N$ limit with
the parameters $\tilde{m} \equiv m N$
and $\tilde{\mu} \equiv \mu \sqrt{N} $ fixed \cite{Osborn:2005ss},
which is called the microscopic limit.
In this section we use $\tilde{m}$ and $\tilde{\mu}$ just to
make it easier to compare our results with 
those in ref.~\cite{Mollgaard:2013qra}.
When we discuss the generalized Banks-Casher relation in 
section \ref{sec:discussion}, however, we have to 
take the large-$N$ limit with $m$ and $\mu$ fixed, 
which is different from the microscopic limit.}
Our result obtained without gauge cooling
is consistent with the one obtained in ref.~\cite{Mollgaard:2013qra}.
As was pointed out there, 
the exact result is reproduced only for $\tilde{m} \gtrsim 10$
and the result for $\tilde{m} \lesssim 10$ turns out to be close to
the result of the phase-quenched model, in which 
the fermion determinant is replaced by its absolute value. 
On the other hand, the result obtained with the gauge cooling
using the norm ${\cal N}_1$ agrees well with the exact result
all the way down to $\tilde{m} \sim 1$.
%quite small quark mass. 
The result obtained with the norm ${\cal N}_2$ is equally good
except for the data points at $\tilde{m} \le 1$, which exhibit 
certain deviation.

\begin{figure}[htbp] %%%%%%%%%%%%%%%%%%%%%%%%%%%%%%%%%%%%%%%%%%%%%%%%%
\begin{minipage}[b]{.5\linewidth}
\centering
\includegraphics[width=7cm]{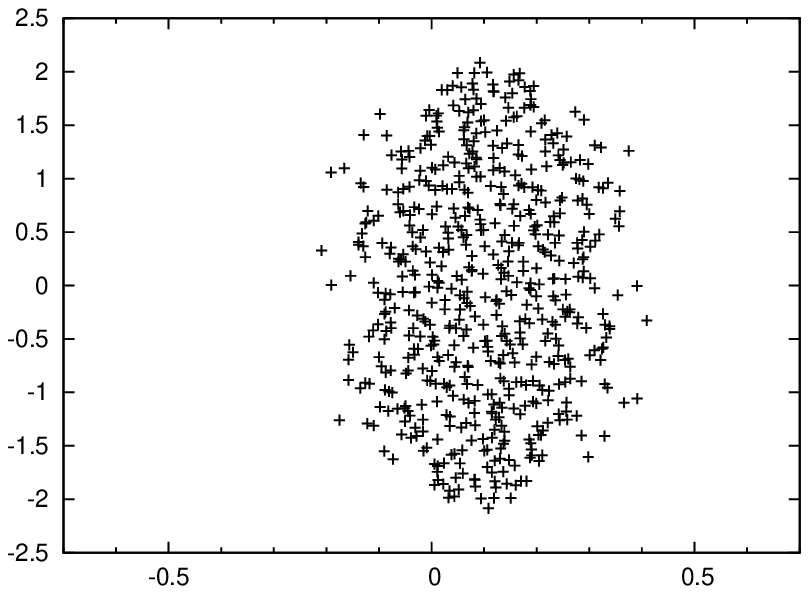}
\subcaption{without gauge cooling}
\end{minipage}
\begin{minipage}[b]{.5\linewidth}
\centering
\includegraphics[width=7cm]{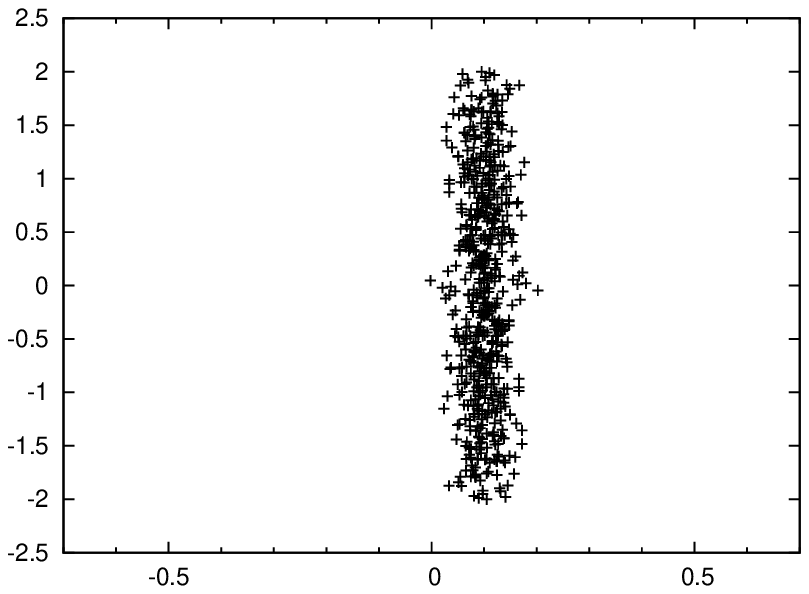}
\subcaption{with gauge cooling using the norm $\mathcal N_1$}
\end{minipage}\\
\begin{minipage}[b]{.5\linewidth}
\centering
\includegraphics[width=7cm]{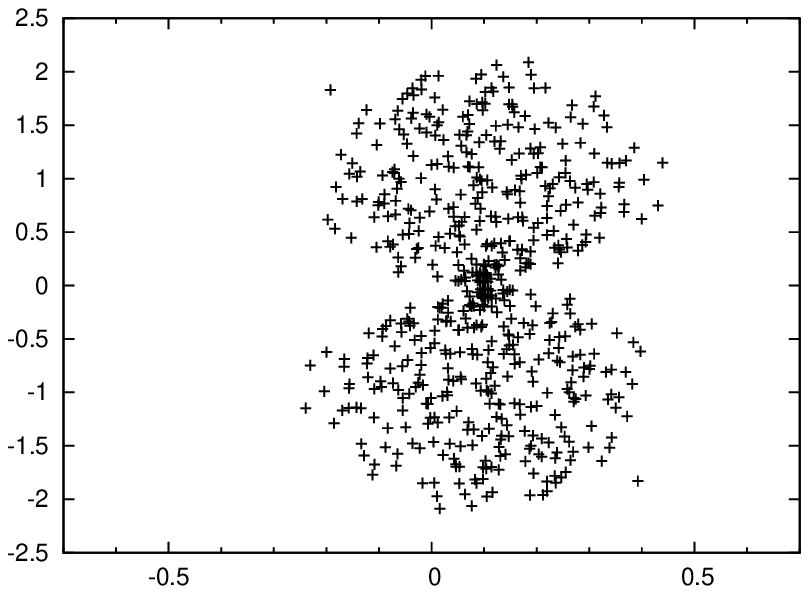}
\subcaption{with gauge cooling using the norm $\mathcal N_2$}
\end{minipage}
\begin{minipage}[b]{.5\linewidth}
\centering
\includegraphics[width=7cm]{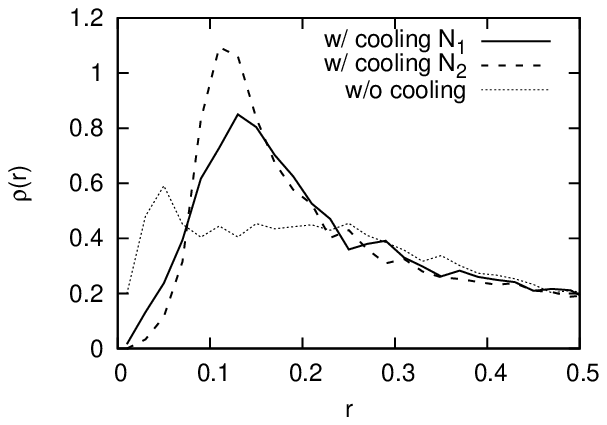}
\subcaption{radial distribution}
\end{minipage}
 \caption{The scattered plots of the eigenvalues of $D+m$ 
obtained at $\tilde{m} \equiv m N = 3$
for the three cases: 
(a) without gauge cooling,
(b) with the gauge cooling using the norm $\mathcal N_1$ and 
(c) with the gauge cooling using the norm $\mathcal N_2$.
The other parameters of the cRMT are the same as
in Fig.~\ref{Fig:2015Jul29Fig1}.
%The results obtained by the CLM for the cRMT with
%$\nu=0$, $N_{\rm f}=2$, $N=30$ and $\sqrt{N} \mu = 2$
%with or without the gauge cooling.
The plot (d) shows the results for the radial distribution
defined by eq.~(\ref{def-radial-dis})
in the three cases.}
 \label{Fig:eigenvalue distribution}
\end{figure}%%%%%%%%%%%%%%%%%%%%%%%%%%%%%%%%%%%%%%%%%%%%%%%%%%%%%%%%%%

The above results suggest that 
the gauge cooling with the new types of norm
can solve the singular-drift problem of the CLM,
which occurs at small quark mass.
In order to see it more directly,
we present in Fig.~\ref{Fig:eigenvalue distribution} (a)-(c)
the scattered plots of the eigenvalues of $D+m$ for $\tilde{m} \equiv m N = 3$
%We find that the eigenvalue distribution 
obtained from the last 10 configurations separated by the interval of 200 steps.
Note that the eigenvalues $(\pm \lambda)+m$ appear in pairs
in all the three cases
for the reason given below eq.~(\ref{anticommuting}).
In the case without gauge cooling,
the eigenvalue distribution covers 
the singularity at the origin, which 
implies that the singular-drift problem occurs.
However, the eigenvalue distribution is changed drastically
by the gauge cooling with the new types of norm.
The norm ${\cal N}_1$ makes the eigenvalue distribution
narrower in the real direction, whereas
the norm $\mathcal{N}_2$ makes the eigenvalues
repelled
%excluded
from the domain near the origin. 
Thus, we find that the gauge cooling with the new types of norm indeed
has an effect of removing the eigenvalues near the singularity.
In order to see this effect more quantitatively,
%how the eigenvalues of $D+m$ near the singularity
%are avoided by the gauge cooling,
let us consider the radial distribution of eigenvalues \cite{Nishimura:2015pba}
defined by 
\begin{align}
\rho(r)=\frac{1}{2\pi r}\int dxdy \, \rho^{\rm (CL)}(x,y)
\, \delta(\sqrt{(x+m)^2+y^2}-r) \ ,
\label{def-radial-dis}
\end{align}
where $\rho^{\rm (CL)}(x,y)$ is the eigenvalue distribution of 
the Dirac operator $D$ obtained during the complex Langevin simulation
with $x$ and $y$ representing the real part and the imaginary part of
the eigenvalue, respectively.
(See (\ref{ev dist}) for the precise definition of $\rho^{\rm (CL)}(x,y)$.)
%whose precise definition will be given in the next section.
In Fig.~\ref{Fig:eigenvalue distribution} (d),
we plot the results for the radial distribution in the three cases discussed above.
We find that the radial distribution is strongly suppressed near the singularity
by the effect of the gauge cooling.
%However, this plot alone cannot explain
%our observation from Figure \ref{Fig:2015Jul29Fig1}
%that the norm $\mathcal{N}_1$ works slightly better than the norm $\mathcal{N}_2$.

%%%%%%%%%%%%%%%%

%So far we have shown the results obtained 
%for the fixed chemical potential, $\tilde \mu =2$.

\begin{figure}[htbp] %%%%%%%%%%%%%%%%%%%%%%%%%%%%%%%%%%%%%%%%%%%%%%%%%
\begin{center}
\includegraphics[width=8cm]{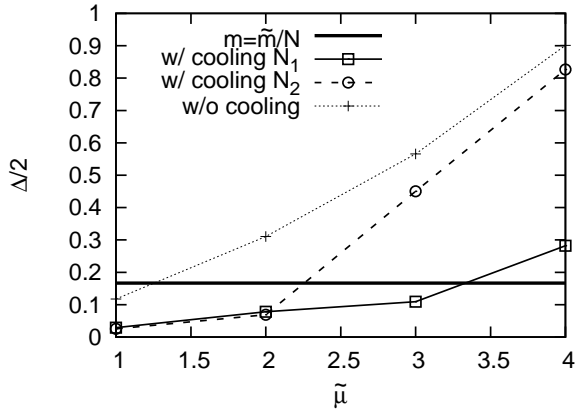}
\caption{The quantity $\Delta/2$
is plotted against $\tilde \mu \equiv \mu \sqrt{N}$
for $\nu=0$, $N_{\rm f}=2$, $N=30$ and $\tilde{m} \equiv m N =5$
with or without gauge cooling.
($\Delta$ represents the width of the eigenvalue distribution
of the Dirac operator $D$ on the real axis.)
The horizontal line represents $m= 1/6$.
In the region $\Delta/2 < m = 1/6$, the singular-drift problem is avoided
and the CLM works.
}
\label{Fig:D_width}
\end{center}
\end{figure}

%A natural question to ask then is: 
Let us then discuss what happens if we increase the chemical potential $\mu$ further.
The exact result for the cRMT is known to be independent of $\mu$
as we mentioned below (\ref{chiral}).
%In the absence of 
When the gauge cooling is not performed,
the eigenvalue distribution of $D+m$ measured during the complex Langevin simulation
tends to become wider in the real direction for increasing $\mu$.
The question is whether the gauge cooling with the new types of norm
can remove the eigenvalues of $D+m$ near the origin even in that case.
%the distribution starts to cover the origin at certain $\tilde \mu$ and 
%causes the complex Langevin simulation to fail.
In order to answer this question,
we measure the width $\Delta$
of the eigenvalue distribution of $D$ on the real axis.
%When $\Delta/2 > m$, the eigenvalue distribution of $D+m$ 
%covers the origin and the CLM fails.
As long as $\Delta/2 < m$, the eigenvalues of $D+m$ do not appear
near the origin and the CLM works. 
%For instance, $\Delta=0.15$ and $\Delta=0.6$ for the case 
%with and without gauge cooling.
In Fig.~\ref{Fig:D_width} we plot $\Delta/2$
obtained by the CLM with $\tilde m =5$ and $\tilde \mu=1,2,3,4$ 
with or without gauge cooling.
We find that the CLM without gauge cooling works only 
%up to $\tilde \mu \sim 1.2$,
for $\tilde \mu \lesssim 1.2$,
whereas the gauge cooling with the norm $\mathcal{N}_1$ and $\mathcal{N}_2$
makes the CLM work 
%up to $\tilde \mu \sim 3.3$ and $\tilde \mu \sim 2.2$, respectively.
for $\tilde \mu \lesssim 3.3$ and $\tilde \mu \lesssim 2.2$, respectively.
We should recall, however, that the norm $\mathcal{N}_2$ has two parameters
$\xi$ and $n_{\rm ev}$, which can be optimized to make the gauge cooling more
efficient.
%It is possible that the norm $\mathcal{N}_2$ may become more effective
%by the optimization.
% as the norm $\mathcal{N}_1$ 
Also, the range of applicability may be further enlarged
by increasing the number of gauge cooling procedures after each Langevin step
or by using a smaller Langevin step-size.

%%%%%%%%%%%%%%%%%%%%%%%%%%%%%%%%%%%%%%%%%%%%%%%%%%%%%%%%%%

\section{The generalized Banks-Casher relation}
\label{sec:discussion}
%%%%%%%%%%%%%%%%%%%%%%%%%%%%%%%%%%%%%%%%%%%%%%%%%%%

In the previous section, we have
seen that the gauge cooling with the new types of norm
cures the singular-drift problem caused by the eigenvalues of 
$D+m$ near the origin. 
%This is achieved by modifying the eigenvalue distribution of the Dirac operator $D$.
%% The eigenvalue distribution of $D+m$
%% measured during the complex Langevin simulation
%% is modified by the gauge cooling, and the singularity can be avoided
%% by either type of norm although 
%% the parameter region in which 
%% the problem is avoided depends
%% on the choice of the norm.
Interestingly, we find that the eigenvalue distribution of the Dirac operator
measured during the complex Langevin simulation
depends on the choice of the norm even in the parameter region where 
the singularity is avoided.
%In this section, we discuss theoretical issues about the cooling procedure for 
%the Dirac eigenvalues.
This is possible because the eigenvalue distribution 
of the Dirac operator is non-holomorphic, and hence
the distribution measured during the complex Langevin simulation
is not directly related to the 
%eigenvalue density
corresponding quantity
defined in the original path integral.
(Note, for instance, that the former is real non-negative by definition,
whereas the latter is complex in general due to the complex weight.)
This is in contrast to the observable such as
the chiral condensate,
which is holomorphic, and hence
the quantity measured during the complex Langevin simulation
should agree with the corresponding quantity defined in the original path integral.
%% Below, we will derive and confirm
%% a generalized Banks-Casher relation, which 
%% expresses the chiral condensate in the chiral limit
%% in terms of the eigenvalue distribution
%% measured during the complex Langevin simulation.
%% This shows, in particular, that
%% the eigenvalue distribution in the chiral limit has a universal property which
%% does not depend on the choice of the norm used in the gauge cooling procedure
%% as long as the CLM works although the distribution itself depends on it.

In fact, the chiral condensate can be written in terms of
the eigenvalue distribution of the Dirac operator.
This leads to the well-known Banks-Casher relation \cite{Banks:1979yr}
at $\mu=0$, which expresses
the chiral condensate in the chiral limit
in terms of the eigenvalue distribution at the origin.
At $\mu\neq 0$, 
the eigenvalue distribution of the Dirac operator defined
in the path integral becomes complex due to the complex weight,
and the violent oscillation of the complex eigenvalue
distribution in an extended region is responsible for the nonzero
chiral condensate in the chiral limit \cite{Osborn:2005ss}.
On the other hand, 
the eigenvalue distribution of the Dirac operator 
measured in the complex Langevin simulation 
is real non-negative,
and therefore its relation to the chiral condensate 
should be analogous to the original one at $\mu=0$.
However, since the eigenvalue distribution spreads out in the complex plane
due to the violation of the anti-Hermiticity of $D$, 
the accumulation of eigenvalues toward the origin should occur
in the chiral limit in order to reproduce 
the non-zero chiral condensate \cite{Splittorff:2014zca}.
What we have seen 
%The behavior 
in the previous section is qualitatively consistent with this argument.
%When the CLM gives correct
In what follows, we make this argument quantitative
by writing down the generalized Banks-Casher relation, 
which relates the chiral condensate with the diverging behavior
of the eigenvalue distribution at the origin in the chiral limit.
We confirm this relation numerically in the cRMT, and discuss
its implication in the context of this work.

First let us define the eigenvalue distribution of the Dirac operator 
measured in the CLM
%in a general theory 
by
\begin{align}
\rho^{\rm (CL)}(x,y)=
\left\langle \frac{1}{n}\sum_{i=1}^{n}
\delta(x-\Re \lambda_i) \, \delta(y-\Im \lambda_i)\right\rangle_{\rm CL} \ ,
\label{ev dist}
\end{align}
where 
%$V$ is the volume of the system,
$\lambda_i$ are the eigenvalues of the Dirac operator $D$, 
$n$ is the number of the eigenvalues,
and $\langle\cdots \rangle_{\rm CL}$ represents 
an ensemble average over the thermalized configurations
generated by the complex Langevin simulation.
Note that the quantity (\ref{ev dist}) for fixed $x$ and $y$
(before taking the ensemble average)
cannot be a holomorphic function of the 
configuration $(\Phi_k , \Psi_k)$ 
since its imaginary part vanishes identically 
while its real part is not a constant.
%it is real and non-negative., and it 
Therefore, the distribution (\ref{ev dist}) 
does not have to agree with
%is generically different from 
the eigenvalue distribution of the Dirac operator defined 
in the original path integral. 
In fact, the latter distribution is generally
a complex-valued function due to the complex weight of the path integral.

As is done in deriving the original Banks-Casher relation at $\mu=0$,
we express the chiral condensate\footnote{The chiral condensate $\Sigma$
defined by (\ref{chiral}) in the cRMT is $2 N_{\rm f}$ times larger than
$\Sigma$ in eq.~(\ref{cond rho}), which is motivated in a more general context.
We will take this extra factor into account 
in arriving at (\ref{cond rho relation crmt}).
}
using the eigenvalue 
distribution $\rho^{\rm (CL)}(x,y)$ as \cite{Splittorff:2014zca}
%of the Dirac operator as \cite{Splittorff:2014zca}
\begin{align}
%\left\langle\frac{1}{V}\mathrm{Tr}\bar q q\right\rangle_{CL}
\Sigma
&=\left\langle \frac{1}{n}\mathrm{Tr}\frac{1}{D+m}\right\rangle_{\rm CL} 
%\nonumber \\
%&
=\int dxdy \, \frac{\rho^{\rm (CL)}(x,y)}{x+iy+m} \ .
\label{cond rho}
\end{align}
%This part can be extracted by noting the symmetry of the eigenvalue distribution:
Note that even after the complexification of the dynamical variables,
the Dirac operator $D$ satisfies \eqref{anticommuting}, 
which leads to $\rho^{\rm (CL)}(x,y)=\rho^{\rm (CL)}(-x,-y)$. 
Therefore, the eigenvalues with $|x+iy| \gg m$ cancel pairwise
%with each other 
in \eqref{cond rho}, and hence small eigenvalues with
%The chiral condensate is dominated by small eigenvalues, 
$|x+iy|\lesssim m$ are responsible for the nonzero condensate.

Let us introduce the polar coordinates $x = r \cos \theta$
and $y = r \sin \theta$, where $-\infty < r < \infty$ and 
$0 \le \theta < \pi$, and the corresponding distribution
$\tilde{\rho}^{\rm (CL)}(r, \theta)$, which has the symmetry
$\tilde{\rho}^{\rm (CL)}(r,\theta)=\tilde{\rho}^{\rm (CL)}(-r, \theta)$. 
Using this symmetry, we can rewrite 
\eqref{cond rho} with the polar coordinates as
\begin{align}
%\left\langle\frac{1}{V}\mathrm{Tr}\bar q q\right\rangle_{CL}
\Sigma 
&=\frac{1}{2}\int_{-\infty}^{\infty} dr |r| 
\int_0^\pi d\theta \, \tilde{\rho}^{\rm (CL)}(r,\theta)
\left( \frac{1}{re^{i\theta}+m}+\frac{1}{-re^{i\theta}+m}\right) \ .
\label{cond rho polar}
\end{align}
%% \begin{align}
%% \left\langle\frac{1}{N}\bar q q\right\rangle_{CL}
%% &=N_{\rm f}\int_{-\infty}^{\infty} dr |r| \int_0^\pi d\theta \rho^{\rm (CL)}(r,\theta)
%% \left( \frac{1}{re^{i\theta}+m}+\frac{1}{-re^{i\theta}+m}\right),
%% \label{cond rho polar}
%% \end{align}
%$\rho^{\rm (CL)}(r,\theta)=\rho^{\rm (CL)}(r,\theta+\pi)=\rho^{\rm (CL)}(-r,\theta)$, 
%which are obtained from \eqref{sym of rho 2}.
In the $m \rightarrow + 0$ limit, 
the terms in the parentheses can be replaced with 
$2\pi i e^{-i\theta}\delta(r)$.
Therefore, \eqref{cond rho polar} can be evaluated as
%% \begin{align}
%% \left\langle\frac{1}{N}\bar q q\right\rangle_{CL}
%% &=2\pi iN_{\rm f} \lim_{r\to 0}  r \int_0^\pi d\theta \rho^{\rm (CL)}(r,\theta)e^{-i\theta} \nonumber \\
%% &=2\pi N_{\rm f} \lim_{r\to 0}   r \int_0^\pi d\theta \rho^{\rm (CL)}(r,\theta)\sin\theta
%% \label{cond rho polar 2}
%% \end{align}
\begin{align}
%\left\langle\frac{1}{V}\mathrm{Tr} \bar q q\right\rangle_{CL}
\lim_{m\rightarrow 0} \Sigma
&=\pi i \lim_{r\to +0}  r \int_0^\pi d\theta \, e^{-i\theta}
\lim_{m\rightarrow 0}  \tilde{\rho}^{\rm (CL)}(r,\theta)   \ .
\label{cond rho polar 2}
\end{align}
Here and henceforth, the chiral limit represented by $\lim_{m\rightarrow +0}$
is assumed to be taken after taking the thermodynamic limit ($n\rightarrow \infty$).

To proceed further, let us recall that
the chiral condensate (\ref{cond rho})
%computed by the complex Langevin method is ensured to be real. 
obtained by the CLM
%real because of symmetry.
is guaranteed to be real by symmetry.
Note that the complex Langevin equations are invariant under
complex conjugation $(\Phi_k , \Psi_k) \to (\Phi_k ^* , \Psi_k^*)$,
under which the Dirac operator becomes $D \to D^*$.
This ensures that the eigenvalue distribution in the thermal
equilibrium of the Langevin process has the symmetry
$\rho^{\rm (CL)}(x,y)=\rho^{\rm (CL)}(x,-y)$,
from which it follows that (\ref{cond rho}) is real.
In the case of polar coordinates,
the symmetry translates to 
$\tilde{\rho}^{\rm (CL)}(r,\theta)=\tilde{\rho}^{\rm (CL)}(-r, \pi -\theta)$,
which implies that (\ref{cond rho polar 2}) is real and leads
to the generalized Banks-Casher relation
\begin{align}
%\left\langle\frac{1}{V}\mathrm{Tr}\bar q q\right\rangle_{CL}
\lim_{m\rightarrow +0} \Sigma
&=\pi \lim_{r\to +0}   r  
%\left(
\lim_{m\rightarrow +0} \hat \rho^{\rm (CL)}(r) 
%\right) 
\ ,
\label{cond rho relation}
\end{align}
where we have defined
%% \begin{align}
%% %\left\langle\frac{1}{V}\mathrm{Tr} \bar q q\right\rangle_{CL}
%% \Sigma
%% &=\pi  \lim_{r\to 0}   |r| \int_0^\pi d\theta \rho^{\rm (CL)}(r,\theta)\sin\theta \, ,
%% \label{cond rho polar 2-positive}
%% \end{align}
%
\begin{align}
\hat \rho^{\rm (CL)}(r) = 
\int_0^\pi d\theta \sin\theta \, \tilde{\rho}^{\rm (CL)}(r,\theta) \ .
\label{rho hat}
\end{align}
The relation \eqref{cond rho relation} implies
that the quantity $\lim_{m\rightarrow +0} \hat \rho^{\rm (CL)}(r)$ should diverge as 
\begin{align}
\lim_{m\rightarrow +0} \hat \rho^{\rm (CL)}(r) \sim \frac{1}{\pi r}
\lim_{m\rightarrow +0} \Sigma
\quad \quad \quad
\mbox{for~}r \rightarrow +0 \ ,
\label{asymp-rho}
\end{align}
%$\hat \rho^{\rm (CL)}(r) \sim \Sigma/ (\pi r)$ for $r \rightarrow +0$
when $\lim_{m\rightarrow +0} \Sigma$ is nonzero; namely
when the chiral symmetry is spontaneously broken.
%% Hence, the asymptotic behavior of 
%% $\hat \rho^{\rm (CL)}(r)$ for $r \rightarrow 0$ is universal
%% as long as the CLM works,
%% although the eigenvalue distribution $\rho^{\rm (CL)}(x,y)$ itself
%% is not a universal quantity as we emphasized above.

When the singular-drift problem occurs, the chiral condensate
obtained by the CLM becomes smaller than the exact 
result (Fig.~\ref{Fig:2015Jul29Fig1} (Left)).
The gauge cooling with appropriate norm
cures the singular-drift problem by making the eigenvalue distribution
of $D+m$ suppressed strongly near the 
singularity (Fig.~\ref{Fig:eigenvalue distribution} (b) and (c)).
This has an effect of making the eigenvalue distribution of $D$
(not $D+m$) larger near the origin, 
and hence the chiral condensate becomes larger.
Note that the CLM may work for different choices of norm as we have shown
in the cRMT. In that case, the eigenvalue distribution of $D$ 
itself may depend on the choice of norm as we emphasized above,
but the asymptotic behavior (\ref{asymp-rho}) of 
$\hat \rho^{\rm (CL)}(r)$
% for $r \rightarrow 0$
in the chiral limit
is universal 
since it is fixed by the generalized Banks-Casher relation (\ref{cond rho relation}).
%chiral condensate.

\begin{figure}[htbp] %%%%%%%%%%%%%%%%%%%%%%%%%%%%%%%%%%%%%%%%%%%%%%%%%
%\begin{center}
%\includegraphics[width=8cm]{histogram_trho_cN1_n10n20n30.eps}
%\includegraphics[width=8cm,angle=-90]{histogram_trho_cN1_n10n20n30_mono.eps}
%%%this is an old file \includegraphics[width=8cm]{histogram_trho_cN1_n10n20n30_mono.eps}
\begin{minipage}[b]{.5\linewidth}
\centering
\includegraphics[width=7cm]{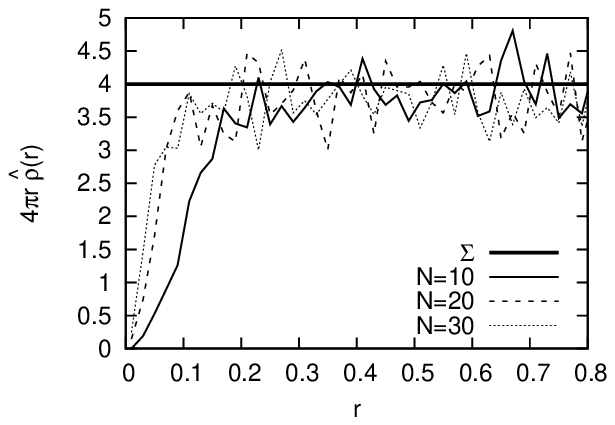}
\subcaption{with gauge cooling using the norm ${\cal N}_1$}
\end{minipage}
\begin{minipage}[b]{.5\linewidth}
\centering
\includegraphics[width=7cm]{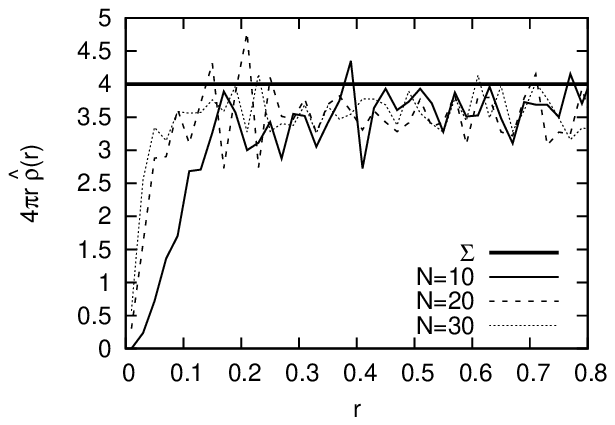}
\subcaption{with gauge cooling using the norm ${\cal N}_2$}
\end{minipage}
\caption{The quantity $2\pi N_{\rm f} r \, \hat \rho^{\rm (CL)}(r)$ 
obtained from the eigenvalue distribution of the Dirac operator
is plotted against $r$ for $N_{\rm f}=2$, $m=0.1$ and $\mu=2/\sqrt{30}$.
The gauge cooling is performed either
with the norm $\mathcal N_1(s)$ or $\mathcal N_2(s)$,
and the results are shown on the left and right, respectively.
%with $r=0.01$ in (\ref{Ntot}) for all the cases.
The horizontal solid line represents the exact result for 
the chiral condensate $\lim_{m\rightarrow +0}\Sigma$ in the cRMT
in the chiral limit.}
\label{Fig:2015Sep15Fig1}
%\end{center}
\end{figure}%%%%%%%%%%%%%%%%%%%%%%%%%%%%%%%%%%%%%%%%%%%%%%%%%%%%%%%%%%

Let us confirm the generalized Banks-Casher relation 
\eqref{cond rho relation} in the cRMT.
Comparing \eqref{chiral} with (\ref{cond rho}), where $n = 2 N$,
we obtain the corresponding relation in the cRMT as
\begin{align}
%\left\langle\frac{1}{N}\bar q q\right\rangle_{CL}
\lim_{m\rightarrow +0}
\Sigma 
&=\lim_{r\to +0}  
%\left( 
2\pi N_{\rm f}\,  r \,  \lim_{m\rightarrow +0} \hat \rho^{\rm (CL)}(r) 
%\right)  
\ .
\label{cond rho relation crmt}
\end{align}
In Fig.~\ref{Fig:2015Sep15Fig1} we plot
the quantity 
$2\pi N_{\rm f}\,  r \,  \hat \rho^{\rm (CL)}(r) $
%in the parenthesis in eq.~(\ref{cond rho relation crmt})
for $N=10,20,30$
with $N_{\rm f}=2$, 
$m=0.1$ and $\mu=2/\sqrt{30}$ (This corresponds to choosing 
$\tilde{m}\equiv mN=3$ and $\tilde{\mu}\equiv \mu \sqrt{N}=2$ for $N=30$).
For $N=10$, we double the statistics 
to reduce the statistical errors.\footnote{We discard the first 40000 steps
for thermalization, and perform 160000 steps for measurements.}
The gauge cooling is performed either
with the norm $\mathcal N_1(s)$ or
$\mathcal N_2(s)$.\footnote{In the case 
of the norm $\mathcal N_1(s)$, we use $s=0$ in (\ref{Ntot}) for $N=30$
as we did in the previous sections,
but we had to use $s=0.01$ for $N=10,20$ to avoid the excursion problem.
We also use $s=0.01$ for $N=10,20,30$ in the case of 
the norm $\mathcal N_2(s)$.}
For either choice of the norm,
we observe a clear plateau, which extends towards the origin as $N$ increases.
This is consistent with our expectation that 
the quantity $\hat \rho^{\rm (CL)}(r)$ diverges as $\sim r^{-1}$
for $r\rightarrow 0$ in the chiral limit.
The height of the plateau, which is almost independent of $N$,
gives an estimate
% $\sim 3.75$
for the r.h.s.\ of \eqref{cond rho relation crmt},
%which is $2\pi N_{\rm f} r\hat{\rho}^{\rm (CL)}(r) 
%for $r\to 0$ and $N\to \infty$, and $m=0.1$, 
which 
%should be compared with 
is close to the exact result $\lim_{m\rightarrow +0} \Sigma = 4$ for the cRMT
in the chiral limit.
%in the $N\to \infty$ and $m\to 0$ limits.
Certain deviation depending on the norm adopted
can be understood as finite $m$ effects.
\section{Summary and discussions}
\label{sec:summary}
%%%%%%%%%%%%%%%%%%%%%%%%%%%%%%%%%%%%%%%%%%%%%%%%%%%

In this paper we proposed a new method
to overcome the singular-drift problem in the CLM, which occurs, for instance, 
in finite density QCD at small quark mass in the confined phase.
%in the CLM can be overcome in the cRMT
%by extending the idea of gauge cooling.
This problem occurs when
the drift term in the complex Langevin equation has singularities,
and the probability of obtaining configurations near the singularities
during the Langevin process is not suppressed sufficiently.
In the case of finite density QCD, 
%% the singularities of the drift term
%% appears due to the zero eigenvalues of the Dirac operator.
the drift term becomes singular when the Dirac operator 
including the mass term
has zero eigenvalues.
%are associated with the small eigenvalues of the Dirac operator.
%in the above parameter region.

Our method is based on the gauge cooling, 
which was originally proposed to overcome
the excursion problem.
Unlike the original proposal, however, we use
new types of norm which are sensitive to the eigenvalue distribution 
of the Dirac operator.
Performing the gauge cooling with the new types of norm 
after each Langevin step, we can remove the eigenvalues close to zero
in such a way that the calculation of gauge invariant observables is not
affected.
We tested the method in the cRMT, which is a simplified model of finite 
density QCD at zero temperature, and confirmed that the method 
allows us to reproduce the exact result even in the small quark mass regime,
which was not accessible without gauge cooling due to the singular-drift problem.

We emphasized that the eigenvalue distribution of the Dirac operator
measured during the Langevin process is not a holomorphic quantity,
and hence it does not have a unique counterpart in the original path integral
with the complex weight.
We derived the generalized Banks-Casher relation, which expresses
the chiral condensate in terms of this non-universal eigenvalue distribution
in the chiral limit, and confirmed it explicitly in the cRMT.
While the eigenvalue distribution for the two successful choices of norm is 
different, the asymptotic behavior at the origin, 
which is linked to the chiral condensate,
is shown to be universal albeit 
with certain finite $N$ and finite $m$ effects.

Our method can be applied to QCD in a straightforward manner.
In particular, the norm (\ref{normtype2}) can be used
for any lattice Dirac operator.
It is technically important here that one only needs low-lying eigenvalues
of a large Hermitian matrix, which can be obtained efficiently 
by the Lanczos method.
On the other hand, the norm (\ref{normtype1}) can be used
for the staggered fermion since the corresponding Dirac operator 
has the property $D(\mu)^\dag = - D(-\mu)$ as in the cRMT,
but it cannot be used for the Wilson fermion.
The generalized Banks-Casher relation can also be used for the staggered fermion
but not for the Wilson fermion.
%% On the other hand, the norm like (\ref{normtype1}) can be used
%% only for the Dirac operator with good chiral property
%% which allows clear separation of the mass term.
%% For instance, it can be used for staggered fermion straightforwardly
%% since the corresponding Dirac operator has the property 
%% $D(\mu)^\dag = - D(-\mu)$ as in cRMT.\footnote{In the case of 
%% the Ginsparg-Wilson Dirac operator, 
%
It would be interesting to see
to what extent our method 
%can cure the singular-drift problem and thereby 
can enlarge the range of applicability of the CLM
in finite density QCD.
%is an important open question.
%nteresting open question.
%to the small quark mass regime in the confined phase of QCD.

%singular-drift problem 

%% One may 
%% On the other hand, 
%% further investigations are necessary for adjustment of some parameters 
%% used in the cooling procedure.

%%%%%%%%%%%%%%%%%%%%%%%%%%%%%%%%%%%%%%%%%%%%%%%%%%
\acknowledgments
%%%%%%%%%%%%%%%%%%%%%%%%%%%%%%%%%%%%%%%%%%%%%%%%%%

The authors would like to thank
%We thank 
H.~Matsufuru and D.~Sexty for valuable discussions.
This work was supported in part by Grant-in-Aid 
for Scientific Research (No.\ 26800154 for K.N.\ and
No.\ 23244057, 16H03988 for J.N.)
%and No.\ 20340048(B) for T.Y.\
from Japan Society for the Promotion of Science.
%This work is supported by JSPS Grants-in-Aid for Scientific Research (Kakenhi) 
%Grants No.\ 00586901 (K.~N.), ....
%K.~N.\ was supported in part by MEXT SPIRE and JICFuS.
K.~N.\ was supported in part by MEXT SPIRE and JICFuS.
He was also supported by YITP for attending the YITP workshop 
``Hadrons and Hadron Interactions in QCD'' (YITP-T-14-03).
Computations were carried out partly on SR16000 at YITP.
%\appendix
%\section{Drift force}
S.~S.\ was supported by the MEXT-Supported Program 
for the Strategic Research Foundation at Private 
Universities ``Topological Science'' (Grant No.~S1511006).

%\bibliographystyle{JHEP}
%\bibliography{ref_CLE}
%\bibliography{ref_cle-rmt}
\providecommand{\href}[2]{#2}\begingroup\raggedright\endgroup

\end{document}